# Dense Stellar Cores in Merger Remnants

J. Christopher Mihos and Lars Hernquist[1]
*Board of Studies in Astronomy and Astrophysics,*
*University of California, Santa Cruz, CA 95064*
*hos@lick.ucsc.edu, lars@lick.ucsc.edu*

## ABSTRACT

We use numerical models which include star formation to analyze the mass profiles of remnants formed by mergers of disk galaxies. During a merger, dissipation in gas and ensuing star formation leave behind a dense stellar core in the remnant. Rather than joining smoothly onto a de Vaucouleurs profile, the starburst population leads to a sharp break in the surface density profile at a few percent the effective radius. While our results are preliminary, the lack of such signatures in most elliptical galaxies suggests that mergers of gas-rich disk galaxies may not have contributed greatly to the population of present-day ellipticals.

*Subject headings:* galaxies:interactions, galaxies:structure, galaxies:elliptical and lenticular, cD, galaxies:evolution, galaxies:starburst

## 1. Introduction

The possibility that many elliptical galaxies formed from mergers of disk galaxies has received considerable attention during the past twenty years (e.g., Toomre 1977; Schweizer 1982, 1990). In this "merger hypothesis," violent relaxation redistributes the luminous mass during a merger to produce a de Vaucouleurs $R^{\frac{1}{4}}$ law characteristic of elliptical galaxies, as verified by simulations of merging disk galaxies (Barnes 1988, 1992; Hernquist 1992, 1993a). The discovery of $R^{\frac{1}{4}}$ law luminosity profiles in a number of merger remnants (Wright et al. 1990; Stanford & Bushouse 1991), including the archetypical NGC 7252 (Schweizer 1982) provides evidence that these objects may in fact evolve to resemble elliptical galaxies. Furthermore, the detection of low surface brightness loops and shells around many ellipticals (Malin & Carter 1980; Schweizer & Seitzer 1988) also supports the merger hypothesis, as these types of structures are thought to occur naturally in disk galaxy mergers (e.g., Hernquist & Spergel 1992; Hibbard & Mihos 1994).

However, several problems remain. Perhaps most serious, as emphasized by Carlberg (1986) and Gunn (1987) and demonstrated quantitatively by Hernquist, Spergel, & Heyl (1993), is the difficulty of explaining the high phase space densities found in the centers of ellipticals. In particular, remnants of disk galaxy mergers possess very diffuse cores, with significant departures from a pure $R^{\frac{1}{4}}$ law even at moderate radii (Hernquist 1992). In principle, this discrepancy can be overcome if the progenitors include dense bulges (Barnes 1992; Hernquist 1993a,b), or by the combined effects of gas dissipation and star formation in the central regions (e.g., Gunn 1987; Kormendy & Sanders 1992). The first option, while certainly plausible, circumvents the question of elliptical galaxy formation by invoking *a priori* small elliptical systems in the progenitor disks. Therefore we turn our attention to the second option, that of increasing the central phase space density in remnants by central starbursts in the merging galaxies.

Many interacting and merging galaxies show evidence for elevated star formation rates, particularly in their central regions (e.g., Bushouse 1987; Kennicutt et al. 1987). Numerical simulations of galaxy interactions show that mergers are effective at driving a significant inflow of gas into the central regions of a merger remnant (Barnes & Hernquist 1991; Hernquist & Barnes 1994), triggering intense starbursts (Mihos, Richstone, & Bothun 1992; Mihos 1992; Mihos & Hernquist 1994ab). CO interferometry of the nuclei of merging galaxies reveals large pools of molecular gas which dominate the mass distribution at small radii (e.g., Sargent & Scoville 1991; Scoville et al. 1991; Wang, Schweizer, & Scoville 1991). The conversion of this gas into stars could, in principle, supply the needed stellar component to raise the central densities to the levels observed in elliptical galaxies. However, whether or not the young and old populations will join "seamlessly" onto an $R^{\frac{1}{4}}$ profile is problematic.

To test this conjecture, we have constructed models of merging disk galaxies which include both gas dynamics and the effects of star formation. Using these models, we examine the spatial distribution of the young population as well as the total mass profile at late times, when the starburst has died out and the remnant is largely relaxed in its inner regions. We find that dissipation in these mergers is too effective; rather than solving the problem of diffuse cores, dissipation and subsequent star formation yield nuclei which are *too dense* compared to normal ellipticals. While preliminary, our results suggest that remnants of gas-rich disk galaxy mergers ought to possess mass profiles which rise sharply in their central few hundred parsecs. The fraction of elliptical galaxies harboring such dense cores, therefore, may help constrain the extent to which disk galaxy mergers have contributed to the present-day elliptical galaxy population.

## 2. Numerical Technique

We model the evolution of merging galaxies using a hybrid $N$-body/hydrodynamics code (TREESPH), including the effects of star formation. The techniques employed in our calculations are described in Hernquist & Katz (1989) and Mihos & Hernquist (1994c), and will be summarized only briefly here. We use a tree algorithm (Barnes & Hut 1986; Hernquist 1987) to calculate the gravitational forces on the particles comprising the galaxies, and smoothed particle hydrodynamics (SPH; see Monaghan 1992) to model the hydrodynamical evolution of the disk gas. A modified "Schmidt law" (Schmidt 1959) is used to model star formation, whereby the star formation rate in a gas particle is calculated by $\dot{M}_i \propto M_i \times \rho_{gas}^{\frac{1}{2}}$. Averaged over volume, this parametrization closely follows a Schmidt law of index $n \sim 1.5$ (Mihos & Hernquist 1994c). When star formation occurs, kinetic energy is imparted to surrounding gas, mimicking the effects of energy deposition via supernovae and stellar winds. The conversion of gas to stars is achieved via the use of hybrid SPH/young star particles. The hydrodynamics of these particles is governed by their gas mass fraction until the particles are depleted of gas, after which they evolve in a purely collisionless manner. In such a manner, the evolution of both the gas and the newly formed starburst population can be followed (see Mihos & Hernquist 1994c for more details).

The model galaxies are constructed as described by Hernquist (1993c), and consist of a self-gravitating exponential disk of mass $M_d = 1$ and radial scale length $r_s = 1$, a spherical isothermal halo of mass $M_h = 5.8$ and core radius $r_c = 1.0$, and, optionally, an oblate central bulge of mass $M_b = 1/3$, major axis scale length $a = 0.2$, and minor axis scale length $c = 0.1$. The total number of collisionless particles used in each component is $N_d$=32768, $N_h$=32678, and $N_b$=8192. The disk gas in each galaxy, comprising 10% of the total disk mass, is initially represented by 16384 hybrid SPH/young star particles. Scaling the model units to match physical units typical of the Milky Way, unit length is 3.5 kpc, unit mass is $5.6 \times 10^{10}$ M$_\odot$, and unit time is $1.3 \times 10^7$ years. The galaxies are initially placed 30 disk scale lengths apart, on parabolic orbits which would have $R_{peri} = 2.5$ on the idealized Keplerian trajectory. The orbital geometry is chosen such that one galaxy is exactly prograde, while the disk of the second galaxy is inclined by 71° from the orbital plane. The two models described here differ only in the internal structure of the galaxies: the first model consists of disk/halo galaxies, while the second model includes optional central bulges.

## 3. Results

The spatial distribution of the starburst population in the remnants will be sensitive to the timescales for merging, gas dissipation, and star formation. If the bulk of the star formation takes place before the gas dissipates angular momentum, then the starburst population will be relatively diffuse, like the older stellar disk. Conversely, if dissipation ends before the starburst begins, the starburst population will be much more compact. Furthermore, a starburst population born early during an interaction can be strongly affected by violent relaxation during the final merger, while populations formed when the merger is nearly complete will experience less violent relaxation.

The evolution of the star formation rate in the mergers is detailed in Mihos & Hernquist (1994b), and we highlight their main points here. In a disk/halo merger, gas is driven into the centers of the galaxies early in the encounter, so that most of the star formation occurs before the final merger. These starbursts deplete the gas in the galaxy centers, yielding a starburst population in each disk which evolves in a collisionless manner thereafter. In contrast, the disk/bulge/halo merger experiences the strongest dissipation and starburst only when the galaxies finally merge, forming a dense starburst population late, which is subject to much less violent relaxation. In both cases, the conversion of individual SPH particles from gas-dominated to collisionless particles occurs rapidly, during the period of strongest dissipation, such that mass-weighted errors in the mass profiles introduced by the use of hybrid particles are minimal.

We calculate the surface density profiles of the merger remnants at a time 500 Myr after the merger, when the inner portions of the remnants have relaxed. Figure 1 and 2 show the projected mass profiles of the remnants viewed down the short axis; the other views are qualitatively similar and are not shown. Note that for similar mass-to-light ratios for the old and young populations, these curves trace the surface *brightness* profiles. At early times the cores will be even more prominent due to the lower mass-to-light ratio of the starburst population, but at later times, the mass-to-light ratios will be more comparable (Mihos & Hernquist 1994d).

In the disk/halo merger model (Figure 1), the total (old + young) mass profile over $\sim 0.1 - 1$ $R_e$ is better fitted by an $R^{\frac{1}{4}}$ profile than are pure stellar disk/halo mergers (Hernquist 1992). However, because of the compactness of the starburst population, rather than smoothly filling in the central deficit of stars, the mass profile shows a kink at $\sim 4\%$ $R_e$, or $R \sim 250$ pc. Outside of this radius, the mass profile still shows some negative curvature due to the dominance of the old disk population in this region. Inside this "break radius," the mass profile steepens significantly, owing to the compact nature of the starburst population. Because the extent of this dense core is similar to the gravitational softening length of the disk particles, the precise slope of the central mass profile is somewhat questionable. However, the general result, that the starburst population produces a clear break in the mass profile of the remnant, seems firm.

Because of the inclusion of a dense bulge component in the disk/bulge/halo mergers, these remnants show a better match to an $R^{\frac{1}{4}}$ profile over the range $0.1 - 1\ R_e$ than do disk/halo mergers (Figure 2; Hernquist 1993a). However, like the disk/halo merger, this remnant also possesses a dense starburst nucleus, resulting in a break in the mass profile at small radii ($\sim 2\%\ R_e$). The smaller break radius can be traced to two effects: the higher projected densities in the bulge component masking the starburst population, and the more compact structure of the burst population. Therefore, while these mergers also result in deviations from $R^{\frac{1}{4}}$ profiles, high spatial resolution may be necessary to unambiguously detect these signatures.

To gauge the detectability of these dense cores in distant galaxies, we created artificial ground based and Hubble Space Telescope WFC2 images of the remnants at varying redshifts, from which "observed" surface density profiles were reconstructed. These profiles show that as more distant remnants are observed, the break in the density profile becomes less pronounced. The break radius moves outward, and the central density excess (over a pure $R^{\frac{1}{4}}$ law) is reduced. However, deviations from an $R^{\frac{1}{4}}$ law should be detectable via ground based imaging out to $\sim 100$ Mpc. The higher spatial resolution offered by HST should easily detect breaks in mass profiles even out to distances of a few hundred Mpc. Recent HST observations of cores of ellipticals show that while many low luminosity ellipticals do have rising profiles, bright ellipticals generally have flat core profiles (Faber 1994); the results presented here thus suggest that mergers of gas-rich disk galaxies are not an attractive mechanism for forming giant ellipticals.

## 4. Discussion

While the results described here hold for all the models run to date, one point of concern in the interpretation of these results is the extrapolation of a star formation methodology based on quiescent disk star formation into the regime of violent merger-induced starbursts. If the star formation rate is more responsive early in the merger than indicated in our models – either through a larger index $n$ for the Schmidt law, or by an increase in the overall efficiency of star formation – the gas could be depleted before the dense cores form. However, such early depletion would not be consistent with the large central gas masses observed in many merging systems (e.g., Scoville et al. 1991). Similarly, a greater amount of starburst energy imparted to the ISM could disperse the dense gas. The ratio of starburst kinetic energy to core binding energy in our models $\sim E_{\rm kin}/(GM_{\rm core}/R_{\rm core}) \sim 10^{-4}$, suggesting that the kinetic energy release would need to be several orders of magnitude greater to significantly affect the cores. In models of isolated disk galaxies, such increased energy input disrupts the vertical structure of the disk (Mihos & Hernquist 1994c). Accordingly, the necessary modifications to our star formation prescription to prevent the formation of dense cores would result in a poor description of the star forming properties of isolated disks. However, given the very different physical conditions between quiescent disk galaxies and starburst systems, such differences may not be unexpected; further simulations varying the efficiency of and feedback from star formation are necessary before our conclusions can be considered robust.

An examination of the *total* mass profiles of nearby merging galaxies also suggests that disk galaxy mergers may indeed form these dense stellar cores. For example, the broad-band light profile of NGC 7252 shows a good match to an $R^{\frac{1}{4}}$ law (Schweizer 1982; Hibbard et al. 1994). However, CO observations reveal $\sim 3.5 \times 10^9\ M_\odot$ of star-forming molecular gas (Wang et al. 1992) dominating the mass distribution in the central kiloparsec. If efficient star formation converts this gas into stars, the luminous mass profile may well eventually show deviations from an $R^{\frac{1}{4}}$ profile. Even more extreme is Arp 220, which also displays an $R^{\frac{1}{4}}$ profile (Wright et al. 1990), yet contains $\sim 2.5 \times 10^{10}\ M_\odot$ of molecular gas in its inner regions (Scoville et al. 1991). Similar molecular gas masses are inferred near the centers of many ultraluminous infrared galaxies (Scoville et al. 1991), all of which are thought to have resulted from mergers. Hence, while these spectacular examples of galaxy mergers may produce objects which look superficially like normal ellipticals, they should also possess dense stellar cores left behind by the induced nuclear starburst.

Can the merger hypothesis be reconciled with the lack of dense stellar cores in most normal ellipticals? Several effects may play a role in hiding these cores or preventing them from forming. In particular:

1. Dust may obscure the central regions to such an extent that the surface brightness profiles may flatten to an $R^{\frac{1}{4}}$ profile. Significant amounts of dust may be present in the centers of merger remnants, due to the evolving starburst population. Dust has been observed in ellipticals in sufficient amounts to modify the light profiles (e.g., Sadler & Gerhard 1985; Ebneter, Djorgovski, & Davis 1988); however, the amount *and* spatial distribution needed to produce a pure $R^{\frac{1}{4}}$ profile would be fortuitous.

2. Star formation may either be incomplete, or else extremely biased towards high mass stars which do not contribute to the light profiles at late times. The former scenario would leave behind significant amounts of cold gas, in excess of the amounts typically detected in normal ellipticals (Bregman, Hogg, & Roberts 1992), while the latter scenario might contribute to the X-ray halos of ellipticals.[1]

3. The progenitor disk galaxies in the merger hypothesis may not resemble the disk galaxies used in our models or those currently merging to form objects such as NGC 7252 or Arp 220. For example, if the gas distribution in the progenitors is flat with radius, less gas will flow to the center of the merger remnant, resulting in less massive cores. Again, however, this alternative may require "tuning" of parameters to yield perfect $R^{\frac{1}{4}}$ profiles in the merger remnants.

4. The physical conditions of the gas in the central regions may be significantly different than described by our model. In particular, increased gas pressure due to magnetic fields (Spergel & Blitz 1992) or high cosmic ray fluxes (Suchkov, Allen, & Heckman 1993), or starburst energy input greatly in excess of that used in our models, may halt the inflowing gas, producing a more diffuse starburst. Previous models of satellite galaxy mergers suggest our results are robust against modest changes in the equation of state of the gas, as order-of-magnitude increases in the central gas pressure are necessary to strongly influence the gas inflow.

5. The inflowing gas may fuel AGN activity rather than starbursts (e.g., Sanders et al. 1988), or the dense stellar core may collapse to form a supermassive black hole (Weedman 1983).

Many of the above alternatives seem plausible; therefore it is premature to declare the merger hypothesis dead. Certainly, disk galaxy mergers are producing some fraction of elliptical galaxies, as evidenced by observations of systems such as NGC 7252. To determine what fraction of ellipticals originate in this fashion, signatures of the merging process must be identified. Dissipation and star formation, once hoped to provide a clean solution to the problem of the central phase space density, may provide one such signature. The models described here indicate that mergers of gas-rich disks could leave behind dense stellar cores, and light profiles poorly fitted by de Vaucouleurs $R^{\frac{1}{4}}$ laws at small radius. The fraction of elliptical galaxies harboring such dense cores may therefore help constrain the role disk galaxy mergers played in the formation of the present-day elliptical galaxy population.

This work was supported in part by the San Diego Supercomputing Center, the Pittsburgh Supercomputing Center, the Alfred P. Sloan Foundation, NASA Theory Grant NAGW–2422, the NSF under Grants AST 90–18526 and ASC 93-18185, and the Presidential Faculty Fellows Program.

---

[1] For objects like Arp 220, complete conversion of the molecular gas into hot X-ray emitting gas would result in ten times the amount of hot gas typically observed in ellipticals ($\sim$ a few $\times 10^9$ M$_\odot$, Bregman et al. 1992). Thus, the ultraluminous IR galaxies may not evolve into "normal" ellipticals on this evolutionary path.

---

This 2-column preprint was prepared with the AAS LaTeX macros v3.0.

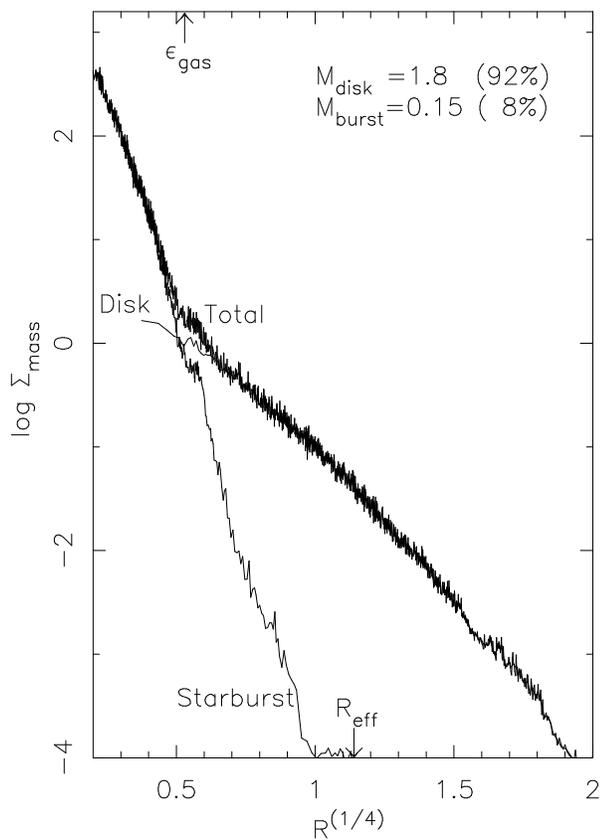 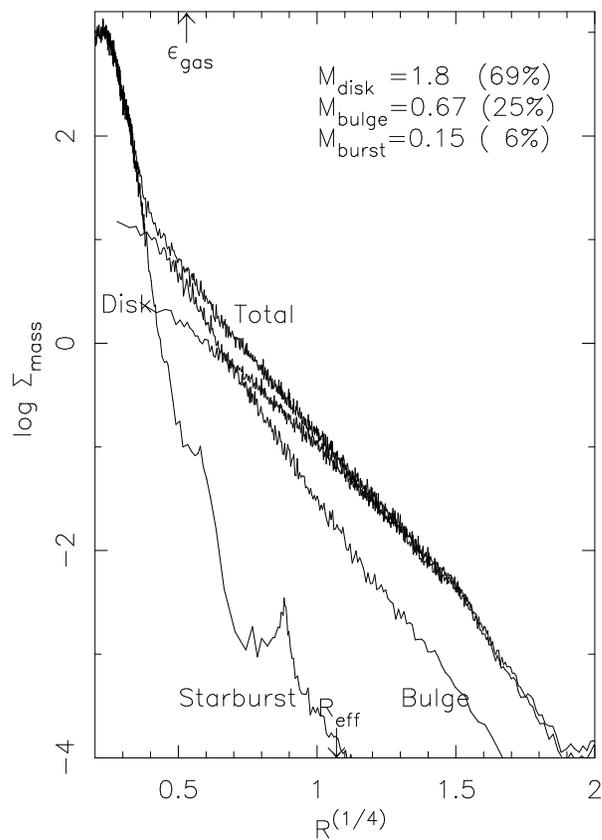

Fig. 1.— Luminous mass profiles of disk/halo galaxy merger remnant. The amount of mass in each component is shown, along with the effective radius of the remnant. The gravitational softening length of the disk and gas particles is also marked.

Fig. 2.— Luminous mass profiles of disk/bulge/halo galaxy merger remnant. The amount of mass in each component is shown, along with the effective radius of the remnant. The gravitational softening length of the disk and gas particles is also marked.